\documentclass[aps,pra,amsmath,amssymb, twocolumn, showpacs,longbibliography]{revtex4-1}

\usepackage{graphicx} 
\usepackage{dcolumn}
\usepackage{bm}
\usepackage{amsmath}
\usepackage{amssymb}
\usepackage{color}
\usepackage{bbold}
\usepackage{bm}
\usepackage{mwe}
\usepackage[normalem]{ulem} 
\usepackage{placeins}

\newcommand{\out}[1]{{\color{black} #1}}

\newcommand{\new}[1]{{\color{black} #1}}

\newcommand{\bea}{\begin{eqnarray}}  
\newcommand{\eea}{\end{eqnarray}}
\newcommand{\ben}{\begin{enumerate}}
\newcommand{\een}{\end{enumerate}}
\newcommand{\be}{\begin{equation}}
\newcommand{\ee}{\end{equation}}

\renewcommand{\vec}[1]{{\boldsymbol #1}}
\newcommand{\cl}[1]{\hat{\mathcal{#1}}}

\newcommand{\ave}[1]{\langle #1 \rangle}

\newcommand{\tr}{\text{Tr}}

\newcommand{\llangle}[1][]{\savebox{\@brx}{\(\m@th{#1\langle}\)}%
  \mathopen{\copy\@brx\kern-0.5\wd\@brx\usebox{\@brx}}}
\newcommand{\rrangle}[1][]{\savebox{\@brx}{\(\m@th{#1\rangle}\)}%
  \mathclose{\copy\@brx\kern-0.5\wd\@brx\usebox{\@brx}}}

\begin{document}
\title{Critical behavior near the many-body localization transition in driven open systems}

\author{Zala Lenar\v{c}i\v{c}$^{1}$}
\thanks{These two authors contributed equally}
\author{Ori Alberton$^{2}$}
\thanks{These two authors contributed equally}
\author{Achim Rosch$^2$}
\author{Ehud Altman$^1$}
\affiliation{$^1$Department of Physics, University of California, Berkeley, California 94720, USA}
\affiliation{$^2$Institute for Theoretical Physics, University of Cologne, D-50937 Cologne, Germany}


\begin{abstract} 
Coupling a many-body localized system to a thermal bath breaks local conservation laws and washes out signatures of localization. When the bath is non-thermal or when the system is also weakly driven, local conserved quantities acquire a highly non-thermal stationary value. We demonstrate how this property can be used to study the many-body localization phase transition in weakly open systems. Here, the strength of the coupling to the non-thermal baths plays a similar role as a finite temperature in a $T=0$ quantum phase transition. By tuning this parameter, we can detect key features of the MBL transition: the divergence of the dynamical exponent due to Griffiths effects in one dimension and the critical disorder strength. We apply these ideas to study the MBL critical point numerically. The possibility to observe critical signatures of the MBL transition in an open system allows for new numerical approaches that overcome the limitations of exact diagonalization studies. Here we propose a scalable numerical scheme to study the MBL critical point using matrix-product operator solution to the Lindblad equation.
\end{abstract}

\maketitle

Many-body localization is a state of interacting quantum systems, which fail to thermalize subject to their intrinsic dynamics due to the effect of strong disorder~\cite{Basko06,Mirlin05,abanin19}. A pertinent question, currently under intense theoretical study and debate, concerns the nature of the phase transition between the ergodic and localized phases. The transition represents a new class of dynamical quantum phase transitions, which involves a fundamental change of the entanglement structure in all, or at least many, of the eigenstates. 
The many-body localization transition is sharp only if the system is completely isolated, which imposes severe limitations on the ability to study it using standard theoretical, numerical, and experimental approaches. 
The requirement of a closed system appears to preclude experiments with solid state materials due to coupling to a phonon bath. Even in experiments with ultra-cold atoms and ion traps, which are usually considered to be exquisitely isolated, signatures of many-body localization are visibly polluted by extrinsic decay processes~\cite{Bloch15,luschen17a,lukin19,Monroe16}. 
\out{Numerical experiments are also severely limited.} Because of the need to address closed system dynamics, these have been mostly restricted to exact diagonalization (ED) of very small systems~\cite{luitz15,Oganesyan2007,Pal2010,Kjall2014,Khemani2017,Serbyn2016, Vasseur2015}. There is increasing evidence that such simulations are overwhelmed by transient finite-size effects that supersede the critical scaling behavior\cite{suntajs19,abanin19a,panda19}.

We propose to bypass the limitations posed by closed systems by studying  signatures of the MBL transition in weakly open driven systems. In a previous work \cite{lenarcic18} some of us showed that in the limit of vanishing coupling $\epsilon$ to a bath and concomitantly weak drive strength $\epsilon \theta$, the MBL transition shows up as a singular change in the temperature variations across the sample. On the thermalizing side of the critical point the temperature fluctuations vanish in the limit $\epsilon\to 0$, while they remain finite on the MBL side.
At non-vanishing coupling $\epsilon$ one expects this transition to broaden into a universal crossover governed by the critical point located at $\epsilon\to 0$.
The dissipative coupling $\epsilon$ has a role similar to turning on a nonzero temperature above a $T=0$ quantum phase transition. Studying the leading dependence of the spatial temperature fluctuations on $\epsilon$ in the vicinity of the critical point is analogous to studying the leading dependence of the order parameter on the temperature in a conventional quantum phase transition. Such a measurement allows to determine critical exponents as well as the critical disorder strength. 
Furthermore, the MBL transition in one dimensional systems is thought to be preceded by a thermal Griffiths regime leading to sub-diffusive transport \cite{znidaric10,nandkishore14,agarwal15,johri15,fischer16,znidaric16,medvedyeva16,bordia16,prelovsek16,levi16,luschen17,luitz17,nandkishore17,everest17,marino18,rubio18,schulz18,weiner19,schulz19,mendoza19}. We use  an effective model of the Griffiths phase to show that the leading dependence of the temperature variations on the dissipative coupling $\epsilon$ reveals the continuously varying dynamical exponent $z$.

The open systems framework facilitates a new computational scheme to investigate the MBL transition, while overcoming the limitations of exact diagonalization. We use a truncated matrix-product operator to represent the density matrix of a disordered system described by a Lindblad equation with coupling $\epsilon$ to dissipators. We find a sharp signature of the Griffith regime with a continuously varying dynamical exponent that diverges at the critical point. 
We note the connection to Refs.~\cite{znidaric16,schulz18,mendoza19}, where Griffiths exponents have been computed numerically for a  spin-chain coupled to Lindblad operators placed at the two ends of the chain to drive a steady state current. Because we study coupling to bulk Lindblad operators, the calculation can converge faster, allowing to access parameter regimes much closer to the MBL transition.

{\it Hydrodynamic description--}
Consider a disordered ergodic system, weakly coupled to a thermal bath with temperature $T_0$ and a drive that heats the system; 
for example, a spin chain, coupled to phonons and driven by light~\cite{lenarcic18}.
Deep in the ergodic phase the system will reach a nearly thermal steady state, with smooth temperature variations determined by a heat flow equation supplemented by sink and source terms due to the coupling to the bath and drive, respectively,
\begin{equation}\label{EqContinuity}
\partial_t e(\vec r) - \nabla \cdot(\kappa(\vec r) \nabla T(\vec r)) = - \epsilon\, g_{1}(\vec r) (T(\vec r)-T_{0}) + \epsilon\,\theta g_2(\vec r) 
\end{equation}
We assumed, for simplicity, that energy is the only conserved quantity in the limit $\epsilon\to 0$.
The disorder in the underlying model is translated to a weak modulation of the conductivity
$\kappa(\vec r)=\bar \kappa +\delta\kappa(\vec r)$ 
and of the couplings $g_{1(2)}(\vec r)=\bar{g} + \delta g_{1(2)}(\vec r)$ to the thermal bath ($g_{1}(\vec r)$) and the drive ($\theta g_{2}(\vec r)$). The temperature profile varies around the mean value, $T(\vec r)=\bar T +
\delta T(\vec r)$, where the mean temperature $\bar T=T_{0}+\theta$ is
determined by the relative strength $\theta$ of the drive compared to the coupling to the bath. 

Linearizing Eq.~\eqref{EqContinuity} in the disorder strength for the steady state gives
$(-\bar\kappa \nabla^{2} 
+\epsilon \bar g)\delta T(\vec r) 
=\epsilon \,\theta \,\delta g(\vec r),
$
where $\delta g(\vec r)= \delta g_2(\vec r)-\delta g_1(\vec r)$. 
This equation is solved for the local temperature variations using the Green's function, 
$\delta T(\vec r)=\epsilon \theta \int d\vec r' G(\vec r-\vec r')  \delta g(\vec r')$, which is in momentum space given by
$\tilde{G}(\vec k)=\left(\bar\kappa\, k^2 +  \bar g\, \epsilon\right)^{-1}$. We can generalize to the sub-diffusive regime
heuristically by using a renormalized Green's function, $G(\vec k)=\left(\bar{\gamma} |k|^z +  \bar  g\, \epsilon\right)^{-1}$, imposing 
dynamical scaling with exponent $z\ge 2$. 
Assuming Gaussian disorder with short range correlations, $\langle \delta g(\vec r)\delta g(\vec r') \rangle \equiv (\delta g)^2 \delta(\vec r-\vec r')$, we find in  $d$ dimensions,
\begin{equation}\label{EqDeltaTG}
\delta T \sim \theta\,  \vert\delta g / \bar{g}\vert \left({\bar{g} /\bar{\gamma}}\right)^{d/2z}
\,\, \epsilon^{d/2z},
\end{equation}
see Suppl. Mat. (SM), \cite{SM}. This approach, however, may not properly account for the effect of rare regions that dominate the transport in the Griffiths regime. Below we examine a minimal model that takes this physics into account.

\noindent {\it Thermal resistor network  --} 
As a minimal model for the Griffiths regime we consider a chain of conducting islands, each characterized by its own temperature $T_i$, coupled by links representing insulating regions of size $\ell$. Together with the energy sink and source terms, this leads to rate equations 
\begin{align}\label{EqDisCont}
\partial_t e_i- \Gamma_{i,i+1} (T_{i+1}&-T_i) +  \Gamma_{i-1,i} (T_{i}-T_{i-1})\\
&= - \epsilon g_{1,i} (T_i-T_{0}) + \epsilon \theta g_{2,i}\notag.
\end{align} 
While this equation may look like a simple discretization of Eq.~(\ref{EqContinuity}), there is a crucial difference coming from the probability distribution of the link conductances. An `insulating' link of length $\ell$ has conductance $\Gamma_{\text{ins}}(\ell)=\Gamma_0 e^{-\ell/a}$ with $a$ being a microscopic scale. Close to the critical point we expect the lengths distribution of insulating regions to be $p(\ell) = \frac{1}{\mathcal{N}} e^{-\ell/\xi}$ with $\xi$  the diverging correlation length, leading to the distribution of link conductances $P(\Gamma)\sim\left(\Gamma/\Gamma_0\right)^{\alpha-1}$ with $\alpha=a/\xi\ll1$. In this case the average resistivity of the chain $\langle \Gamma^{-1} \rangle$ diverges, indicating sub-diffusive transport \cite{hulin90,agarwal15}.

Coupling this system to a bath and to an energy source destroys the insulating behavior of the links, adding a channel of conductance through the link with conductivity $\epsilon \kappa_0$. Thus we take the heat conductance through a link to be
$\Gamma = \epsilon\frac{\kappa_0}{\ell} + \Gamma_0 e^{-\ell/a}$, with an implicit cutoff $\ell\ge a$.
Finally we take $g_{1(2),i}=\bar g + \delta g_{1(2),i}$ with $\delta g_{1(2),i}$ drawn from a uniform distribution in the range $[-\delta g_{1(2)},\delta g_{1(2)}]$. 

We solve for the steady state of Eq.~(\ref{EqDisCont}) numerically to obtain temperature profiles and extract the normalized variation of local temperatures
$
{\delta T}/{\bar{T}} ={\sqrt{ \langle\!\langle \textrm{Var} (T_i) \rangle\!\rangle }}/{ \langle\!\langle \mathbb{E} (T_i) \rangle\!\rangle}.	
$
Here $\mathbb{E}$ and $\textrm{Var}$ are the sample mean and variance, while $\langle\!\langle \cdot \rangle\!\rangle$ denotes averaging over disorder realizations. We assumed that the conducting islands are all of similar size.

\begin{figure}[t!]
\center
\includegraphics[width=\linewidth]{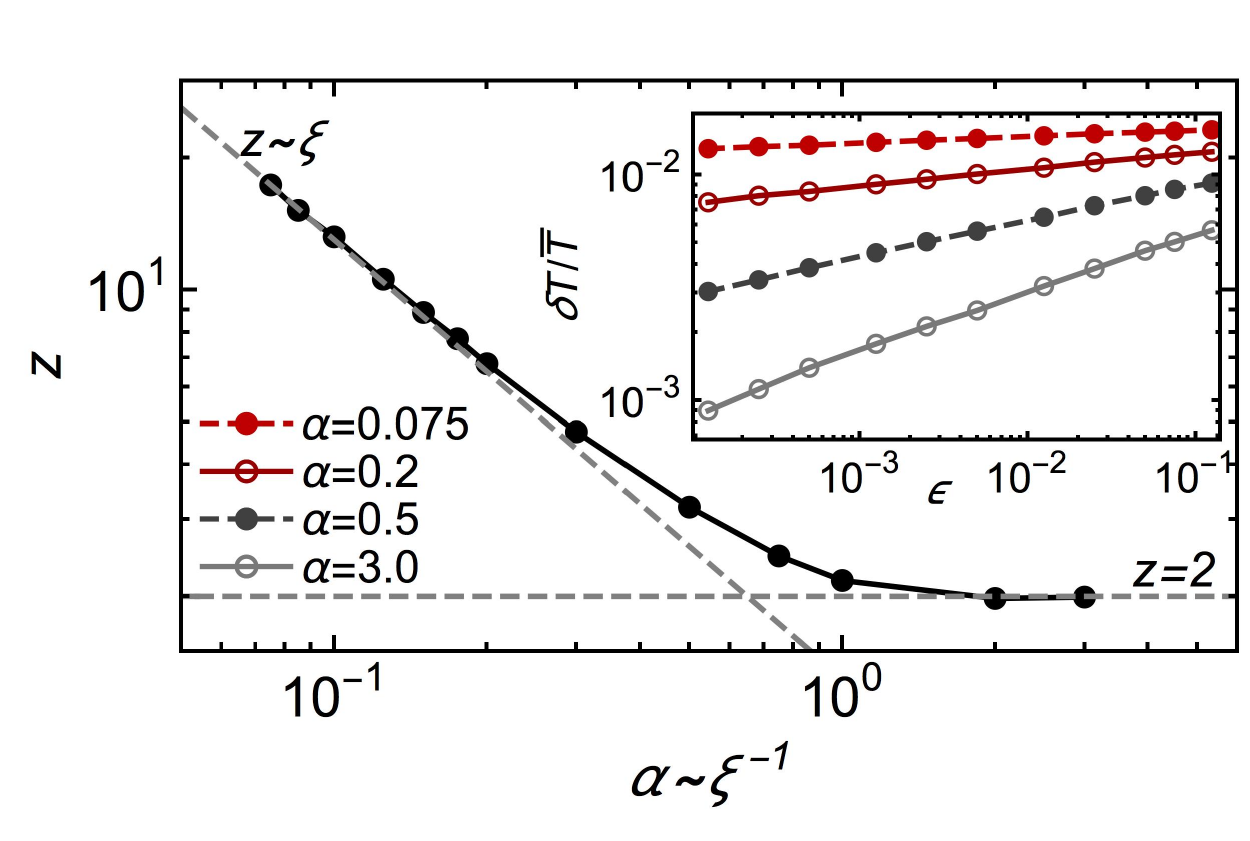}\\
\caption{Fluctuations of the local temperatures, $\delta T/\bar{T}$, are computed from a resistor network model as function of the coupling strength $\epsilon$ to a thermal bath and driving.
For small $\epsilon$, temperature fluctuations are described by $\delta T/\bar{T} \sim \epsilon^{1/2z}$ with $z\sim\alpha^{-1}=\xi/a$ for $0<\alpha<1$, and $z=2$ in the diffusive regime, $\alpha > 1$. 
Parameters: $\kappa_0=1$, $a\Gamma_0=5.0$, $T_0=1$, $\delta g_1=0.05$, $\delta
g_2=0.05$, $\theta= T_0$, $N=1000$, averaged over $M=500$ configurations.
\label{FigRN}
}
\end{figure}

The results for ${\delta T/\bar{T}}$ as a function of $\epsilon$ are shown in 
Fig.~\ref{FigRN}  for different values of $\alpha=a/\xi$. We get $\delta T \sim \epsilon^{1/2z}$ as anticipated in Eq.~\eqref{EqDeltaTG}. 
We see that for $0<\alpha<1$ the dynamical exponent grows with the correlation length as $z \sim 1/\alpha=\xi/a$, whereas for $\alpha>1$ it saturates to $z=2$, as expected for a diffusive system. 
Thus we establish a direct relation between the leading dependence of the temperature fluctuations $\delta T$ on $\epsilon$ and the dynamical exponent $z$, which governs the sub-diffusive behavior in a closed system~\cite{vosk15,potter15}. 

Before proceeding  we comment on the behavior of the thermal resistor network in two dimensions. It is shown in the SM that in this case we have $\delta T\sim \epsilon^{d/4}$, implying $z=2$ for all $\alpha$. This is also the expected dynamical behavior in a two dimensional closed system because any rare region with large resistance can be short circuited by surrounding smaller resistors \cite{gopalakrishnan16a}. 

\noindent{\it Charge transport --} In solid state systems it is usually much easier to measure charge transport than the local temperature profile. It is therefore natural to seek signatures of MBL or the Griffiths regimes in the resistance of a weakly open system. In order to compute how the resistance scales with the external bath or drive coupling $\epsilon$ we consider a charge resistor network described by
\begin{equation}\label{EqChargeNetwork}
\partial_t n_i - \big(\tilde{\Gamma}_{i,i+1} (\mu_{i+1}-\mu_i) -  \tilde{\Gamma}_{i-1,i} (\mu_{i}-\mu_{i-1})\big)=0.
\end{equation} 
Here $\mu_i$ is the electro-chemical potential on island $i$. $\tilde{\Gamma}_{i,j}$ are charge conductances on links, which are distributed exactly as the thermal conductances in Eq.~\eqref{EqDisCont}. Unlike in Eq.~\eqref{EqDisCont}, there are no source or sink terms because the external coupling to the bath and the drive are assumed to conserve charge. In fact, we can consider a system with just a drive or just tunable coupling to phonons. Both give rise to a parallel channel of ohmic conductivity proportional to $\epsilon$ on the insulating links,
$\tilde{\Gamma}(\ell)= \tilde{\Gamma}_0e^{-\ell/a}+\epsilon\sigma_0/\ell.$
Comparing the first and second term we see that the insulating behavior dominates for $\ell< \ell_* \approx a\ln \epsilon^{-1} + a \ln\ln\epsilon^{-1}$, while the bath or drive induced conductance dominates in longer links. To gain analytic insight we calculate the average resistivity of the chain, $\bar{\rho}=\bar{\ell}^{-1} \int d\ell P(\ell)\tilde{\Gamma}(\ell)^{-1}$,
\bea
\bar{\rho}
&\approx& {1\over \bar{\ell}\tilde{\Gamma}_0}\int_a^{\ell_*} d\ell \xi^{-1} e^{\ell (a^{-1}-\xi^{-1})}
\approx  {\alpha \over \bar{\ell} \tilde{\Gamma}_0}\,\left({\epsilon\over \ln \epsilon^{-1}}\right)^{\alpha-1}
\label{EqRbar}
\eea
where $\alpha=a/\xi\sim 1/z $ and $\bar{\ell} = \int \mathrm{d}  \ell P(\ell) \ell
\sim \xi$. 
A numerical solution confirms this result, see SM.
Thus it should be possible to measure the dynamical exponent $z$ by varying the coupling $\epsilon$ via controlled cooling of the phonon bath or by varying the strength of an external drive. 

\noindent{\it Numerical solution of a spin model --} 
Consideration of weakly open driven systems suggests a new approach for investigating the MBL transition numerically.  Here we calculate how the local temperature variations in a spin-chain model change with the coupling to a weak drive that brings the system to a non-thermal steady state. 
The coherent part of the dynamics is governed by the Hamiltonian
\begin{equation}
H=\sum_{i} S_i \cdot S_{i+1} + h (\zeta_i^z S^z_i + \zeta_i^x S^x_i) 
\label{Hmic}
\end{equation}
with open boundary conditions and disorder fields drawn uniformly from the range $\zeta_i^{x,z}\in [-1,1]$. For simplicity we have chosen a model in which energy is the only conserved quantity. 
The MBL transition in the Hamiltonian \eqref{Hmic} has been studied in Ref.~\cite{geraedts17} using exact diagonalization.

To obtain a non-equilibrium steady state we model weak coupling of the system to non-thermal baths within a Lindblad formalism: 
\bea
\dot{\rho}=-i[H,\rho] + \epsilon \sum_{\nu,i} \left(L_i^{\nu} \rho {L_i^{\nu}}^\dagger-\frac{1}{2} \{{L_i^{\nu}}^\dagger L_i^{\nu},\rho\}\right).
\eea
with the Lindblad operators
\begin{align}\label{EqLindblad}
L_i^{\pm,1} &= \frac{S_i^\pm}{2\sqrt{2}} \left(\mathbb{1} - 2 S_{i+1}^z\right) , \
L_i^{\pm,2} = \left(\mathbb{1} - 2S_{i}^z\right) \frac{S_{i+1}^\pm}{2\sqrt{2}}, \notag \\
L_i^{z} &= S_i^z .
\end{align}
The precise choice is not important as long as some of the dissipators are non-Hermitian to ensure a non-trivial steady state, $\rho_{\infty} \neq \mathbb{1}$. In particular, we show in the SM. that a generic Lindblad equation leads to the hydrodynamic Eq.~\eqref{EqContinuity} for the smooth temperature variations. 

The (unique) steady state $\rho_{\infty}$ is obtained by solving the Lindblad time evolution 
using the time-evolving block decimation (TEBD) technique for a vectorized density matrix \cite{verstraete04,zwolak04}. The dephasing term $L_i^{z}$,
Eq.~\eqref{EqLindblad}, is used to ensure that the steady state is sufficiently close to the
identity, so that a bond dimension of $\chi=100$ is adequate to describe a system of $N=20$ sites for
$\epsilon \ge 0.01$. A larger bond dimensions and longer propagation times are needed for smaller $\epsilon$, making computation in these cases more expensive, see SM. 
At fixed $h$, the same set of disorder configurations is used for different values of $\epsilon$, while independent configurations are used at different values of $h$. This procedure helps to determine the exponent $z$ as the $\epsilon$ dependence becomes less affected by the statistical ensemble.
We average over 100 $(h=1,2)$ or 300-500 ($h>2$) disorder configurations.

The goal of the calculation is to obtain the spatial variation of the local temperature for varying values of the dissipative coupling $\epsilon$. To determine the local temperatures $T_i$ we compare the two-site reduced density matrix of the steady state $\rho^{(i,i+1)}_\infty$ with a thermal state \cite{banuls11} by minimizing 
$F[T_i]=\tr\big[ \big(\rho_\infty^{(i,i+1)}-\rho_{\text{th}}^{(i,i+1)}(T_i)\big)^2\big]$ with respect to $T_i$. We chose two sites as the minimal cluster that contains the non-local couplings in the Lindblad equation.

\begin{figure}[t!]
\center
\includegraphics[width=.96\linewidth]{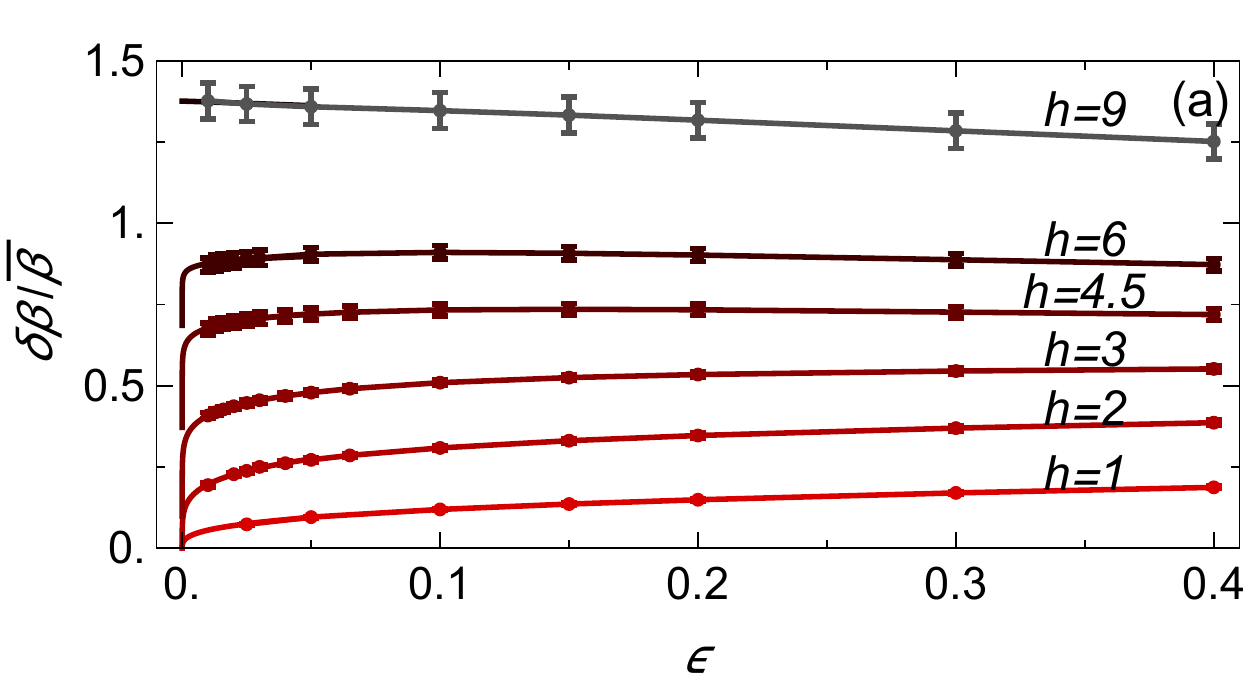}\\
\includegraphics[width=.55\linewidth]{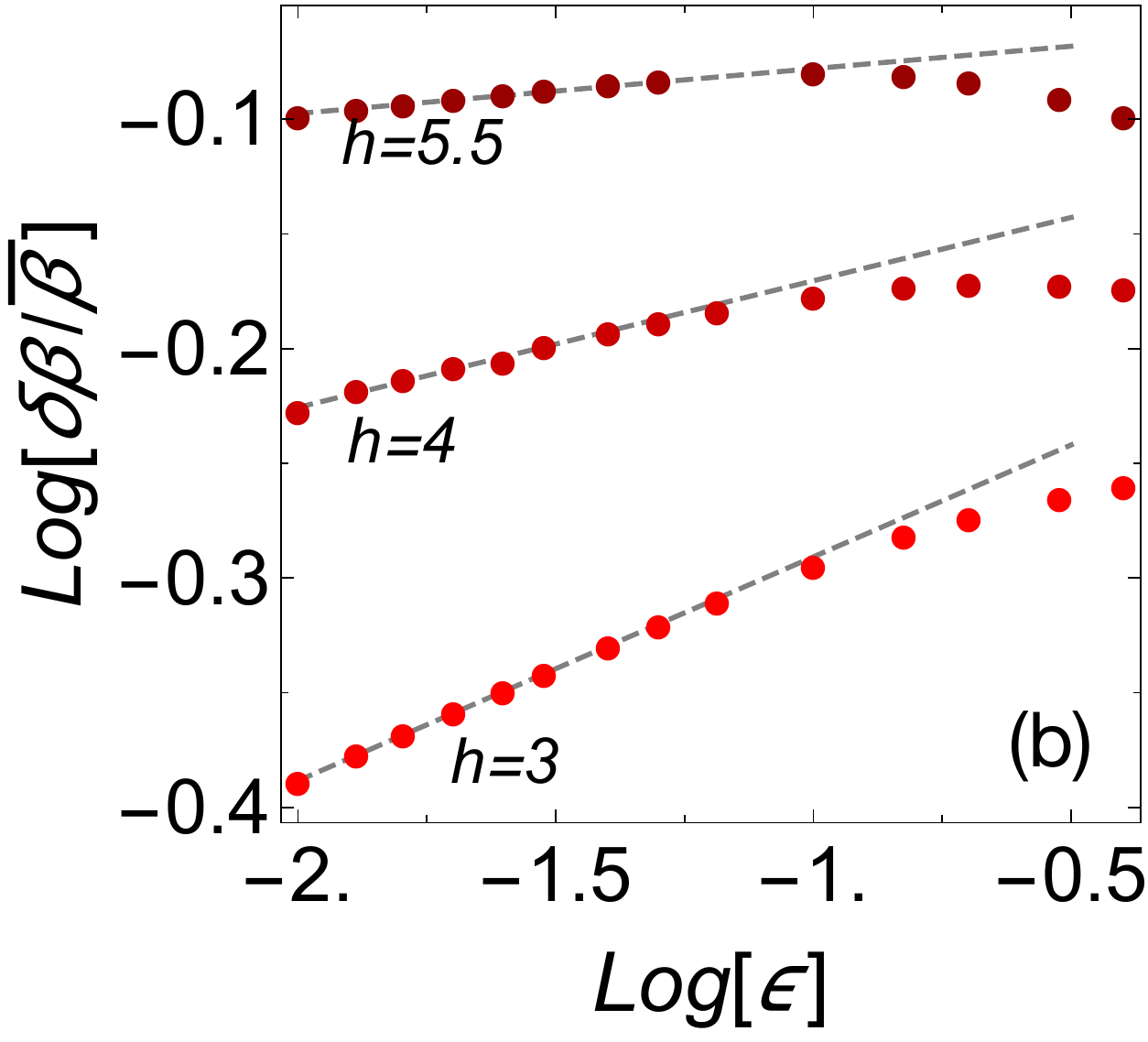}
\includegraphics[width=.41\linewidth]{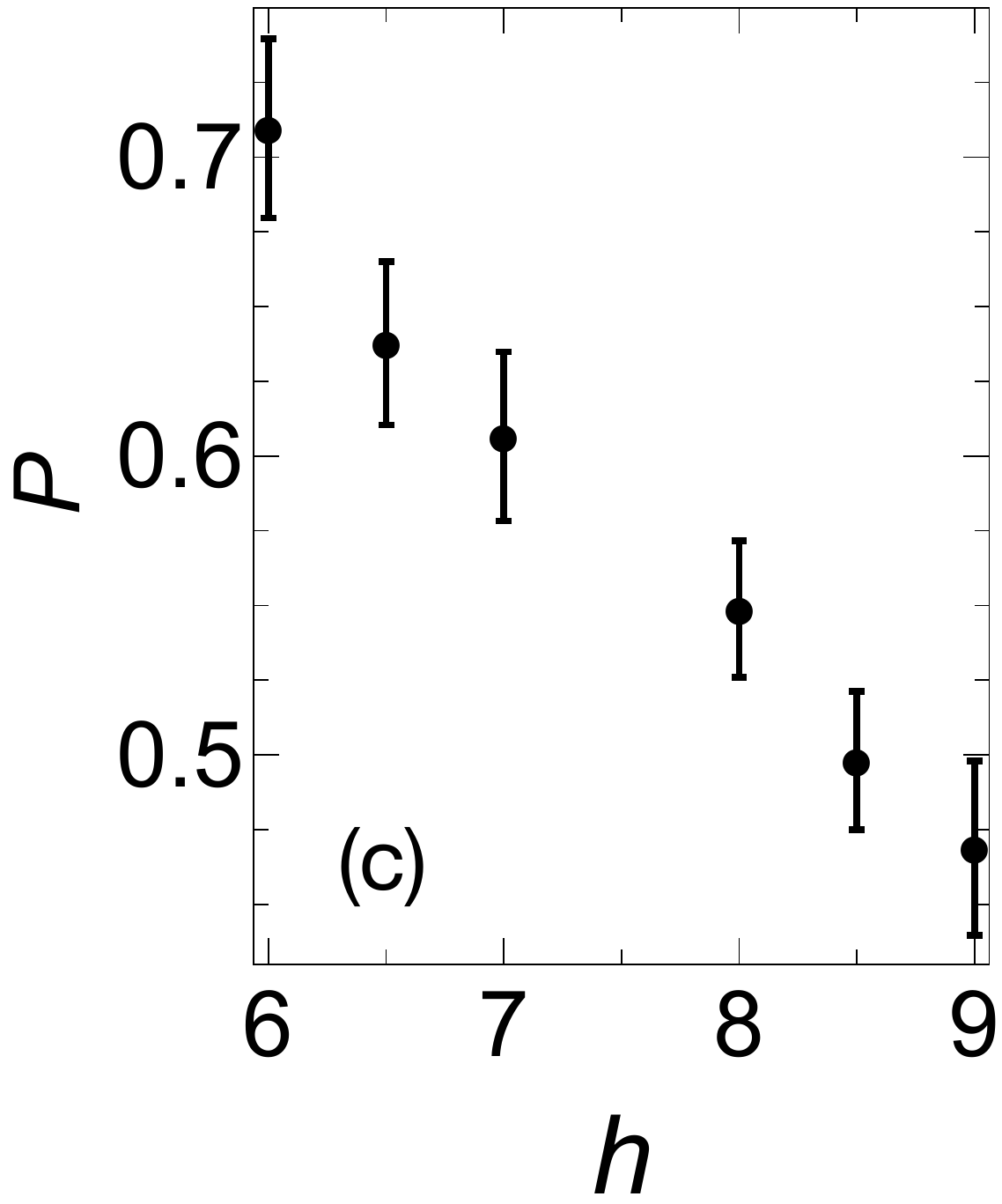}
\caption{{\it Numerical Time Evolving Block Decimation results --}
(a) The fluctuations of the inverse temperature, $\delta \beta/\bar\beta$, show two distinct dependences on $\epsilon$: while they vanish proportionally to $\epsilon^{1/2z}$ on the ergodic side of the phase diagram, they obtain a finite value for $\epsilon \to 0$ with a linear correction in the MBL phase. 
Errorbars show (a correlated) statistical error, while line correspond to the fits.
(b) Fits (dashed lines) to  $\delta \beta/\bar\beta$ are used to obtain $z(h)$ shown in Fig.~\ref{FigZ}.
(c) We estimate $h_c \approx 8.75 \pm 0.5$ from the condition that at $h_c$ the probability $P$ for $\delta\beta/\bar\beta$ to have a positive slope at smallest $\epsilon$ equals $P=0.5$. That is, at $h_c$ the sign of the slope is undetermined.
\label{FigOrdParEps}
}
\end{figure}

\begin{figure}[t!]
\center
\includegraphics[width=\linewidth]{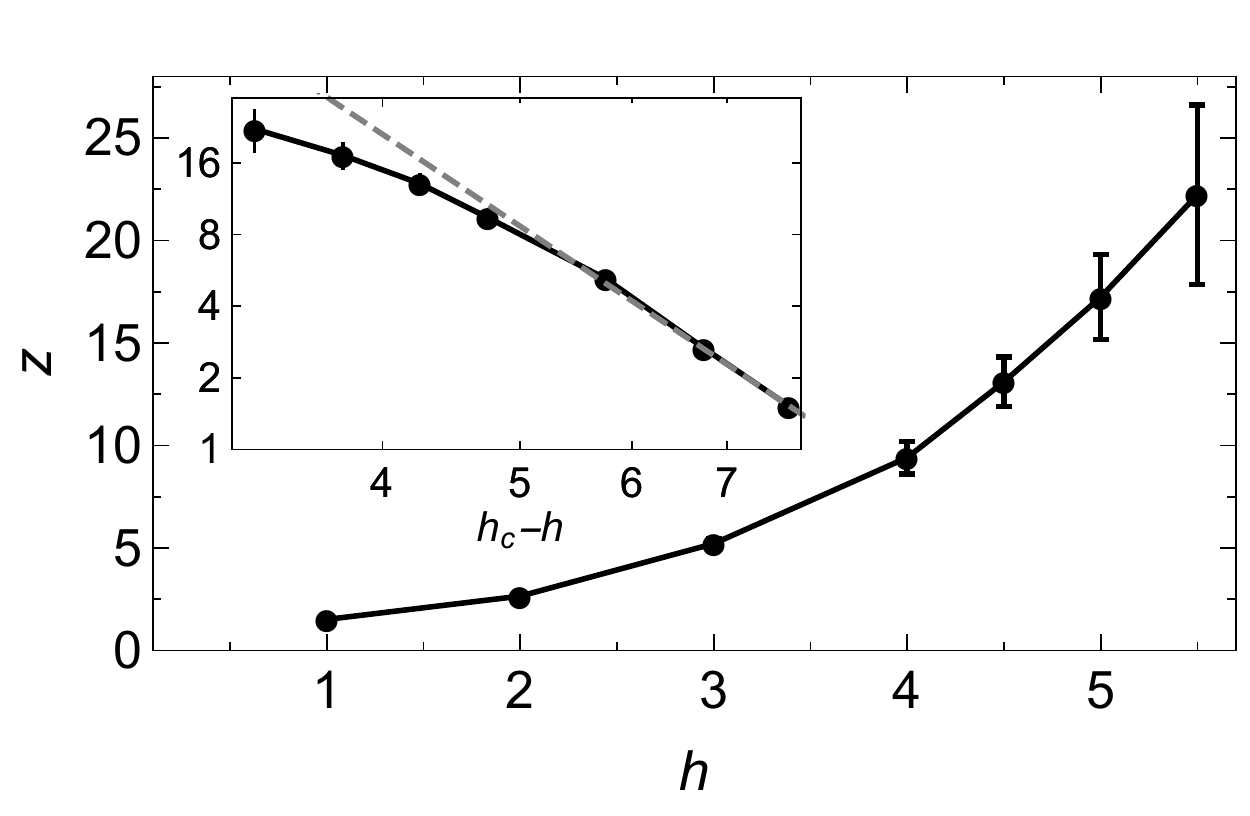}
\caption{The dynamical exponent $z$ increases sharply upon approaching the MBL phase transition. The apparent saturation is due to limitation to $\epsilon \ge 0.01$. 
Inset: $z$ as function of $h_c-h$ on log-log scale assuming $h_c=8.75$.    Using $h_c=8.75\pm0.5$, our results are consistent with $z\sim(h-h_c)^{-\nu}$ with critical exponent $\nu=4\pm0.9$ (dashed line).
\label{FigZ}
}
\end{figure}

The inverse temperature  variations $\delta\beta/\bar{\beta}$, obtained numerically as a function of $\epsilon$, are shown in Fig. \ref{FigOrdParEps} for a range of disorder strengths. We observe different $\epsilon$ dependence in the MBL and ergodic phase, namely
\begin{align}\label{EqOrdParEps}
\frac{\delta \beta}{\bar \beta}(\epsilon) \sim \left\{ 
\begin{array}{ll}
\epsilon^{1/2z}, &h<h_c,\\
\frac{\delta \beta}{\bar \beta}\big|_{\epsilon \to 0} - b \, \epsilon + \mathcal{O}(\epsilon^2),  & h\ge h_c
\end{array}
 \right. .
 \end{align}
In the MBL phase we see temperature variations of order one even in the limit $\epsilon\to 0$ as predicted in Ref. [\onlinecite{lenarcic18}]. At finite $\epsilon$ we expect an analytic dependence on $\epsilon$ due to the local nature of the MBL phase.
In the thermal regime, the temperature variations are expected to vanish in the limit $\epsilon\to 0$ \cite{lenarcic18}. We see an increase of the temperature variations with $\epsilon$ that fits well with the expected non-analytic behavior $\delta\beta/\bar{\beta}\sim \epsilon^{1/2z}$ at small values of $\epsilon$, see Fig.~\ref{FigOrdParEps}(b). The fitted dynamical exponent $z$, shown in  Fig.~\ref{FigZ}, changes continuously with disorder strength, growing rapidly as the MBL transition is approached. Error bars in Fig.~\ref{FigOrdParEps}c and Fig.~\ref{FigZ} were obtained using bootstrap and jackknife resampling, respectively. The usage of a resampling methods for error estimates is necessary 
because statistical errors for different $\epsilon$ at fixed $h$ are strongly correlated in our setup.
As discussed above, the dynamical exponent is expected to diverge together with the correlation length $\xi$ at the MBL critical point. The apparent saturation of $z$ is an artifact of the fit procedure and not a finite-size effect; it is impossible to fit a small $\gamma$ to the function $\epsilon^\gamma$ for realistic values of $\epsilon\gtrsim 0.01$. Thus the minimal value of $\epsilon$ limits the accuracy by which we can determine the critical disorder strength and the critical exponents.
See SM for a systematic finite-size analysis and other numerical aspects.

We obtain a lower bound on the critical disorder strength $h_c$ by recording the fraction $P$ of disorder realizations showing $\delta \beta/\bar{\beta}$ increasing with $\epsilon$ near $\epsilon = 0.01$, Fig.~\ref{FigOrdParEps}(c). 
The estimation of $h_c$ is also limited by the minimal $\epsilon=0.01$ as $h_c$ may increase somewhat if we use a smaller $\epsilon$.
From condition $P=0.5$ we estimate $h_c \geq 8.75\pm 0.5$ for $N=20$,  higher than the value $2<h_c<7$ estimated from an ED study of the same model \cite{geraedts17}. This is consistent with recent analyses suggesting that ED results significantly underestimate $h_c$ due to slow convergence of level spacing statistics \cite{suntajs19,abanin19a,panda19}. Broadening of levels by the dissipative coupling appears to resolve these issues.

We extract a correlation length exponent $\nu$ from the divergence of $z\sim\xi\sim (h_c-h)^{-\nu}$. Using $h_c=8.75\pm 0.5$, we estimate $\nu\approx 4.0\pm 0.9$, which is 
consistent with the Harris-Chayes bound, $\nu>2/d$ \cite{harris74,chayes86} and in agreement with single parameter scaling fits to renormalization group results \cite{vosk15,potter15}. 
But we caution that when using $\epsilon\ge 0.01$ we can reliably fit an exponent only in the range $h\in[1,4]$, not very close to the critical disorder strength.
Based on our current data we cannot exclude Kosterlitz-Thouless like scaling with $z\sim\xi\sim e^{c/\sqrt{h_c-h}}$ as suggested by recent works \cite{goremykina19,dumitrescu19}.

It is an interesting question how the ``order parameter" $\delta\beta/\bar{\beta}$ behaves at the critical point itself  $h=h_c$ in the limit $\epsilon\to 0$. Our results are consistent with a jump across the transition, but they also leave open the possibility of a logarithmic behavior as $-1/\log\epsilon$, which would allow a continuous change of $\delta\beta/\bar{\beta}$ across the transition.

{\it Discussion --} 
We have demonstrated the advantages of investigating the MBL transition as a function of the coupling strength $\epsilon$ to external non-equilibrium baths. 
In numerical computations, the finite coupling to baths limits the operator entanglement entropy allowing to use powerful matrix-product operator  methods on both sides of the phase transition. We were able to obtain quantitative information on quantum critical properties, including the dynamical exponent $z$, the correlation length exponent $\nu$  and the critical disorder strength $h_c$. 
Moreover, having a weak coupling to the baths seems to regulate the calculation by broadening the many-body energy levels facilitating faster convergence to the asymptotic scaling limit.
We also obtain a lower bound on $h_c$, which is much larger than $h_c$ found by direct analysis of ED results \cite{geraedts17}. This is consistent with recent analyses \cite{suntajs19,abanin19a,panda19}, which suggests that $h_c$ is significantly underestimated by ED studies. 

\new{Our numerical scheme is scalable in terms of the calculated spatial size. The limiting factor, instead, is the timescale imposed by the minimal value of the dissipative coupling $\epsilon$. For the system size to dominate the critical scaling, the time $1/\epsilon$ would have to grow at least exponentially with $L$. This is a fundamental limitation stemming from the exponential dynamical scaling that characterizes the critical point. 
}

Our approach to extract MBL from steady-state measurements in driven open systems is complementary to approaches studying the dynamics after a quench \cite{Bloch15,wei18} and opens the door to more experiments with solids in spite of the coupling to phonons. The non-equilibrium conditions discussed in the main text can be achieved by driving the system externally either with light, or with a bias voltage.
Coupling strength to the environment would be tuned, e.g., by modulating the phonon temperature and the power of the light source.
Local temperatures variations can be measured, for example by comparing the Stokes and Antistokes response in a local (tip-enhanced) Raman spectroscopy experiment \cite{anderson00,sternbach17}. Another possibility would be a simpler resistivity measurement, where the dependence of the resistivity on the strength of coupling to phonons can also provide crucial information on the MBL transition. 

\begin{acknowledgments}
Our TEBD code was written in Julia \cite{bezanson17}, relying on the TensorOperations.jl package.
We acknowledge useful discussions with V. Bulchandani, Y. Werman, M.
\v{Z}nidari\v{c}. We acknowledge in part funding from Gordon and Betty
Moore Foundation's EPIC initiative, Grant GBMF4545 and from the European
Research Council (ERC) synergy UQUAM project (EA,ZL), from the ERC under the Horizon 2020 research and innovation program,
Grant Agreement No. 647434 (DOQS) (OA), and from the German Science Foundation under CRC 1238 (project C04) (AR). 
We furthermore thank the Regional Computing Center of the University of Cologne (RRZK) for providing computing time on the DFG-funded High Performance Computing (HPC) system CHEOPS as well as support.
\end{acknowledgments}


\clearpage
\section*{Supplementary material}

\new{
\subsection*{Fractional diffusion}
}
In this section, we use the formalism of fractional calculus \cite{bayin16,kilbas06} in order to
generalize the hydrodynamic Eq.~(1) to describe the
sub- and super-diffusive case. The key idea is to modify the Green’s function
$G(\vec{r})$, describing the response of temperature fluctuations to a random pump and
drive $\delta g(\vec{r})$ (recall that 
$\delta T(\vec{r})=\epsilon\theta\int d\vec{r}'
G(\vec{r}-\vec{r'})\delta g(\vec{r'})$), by replacing the Laplace operator
$\nabla^2$ with the Riesz fractional derivative $\nabla^z$. The fractional
derivative is defined via its Fourier transform $\mathcal{F}$,
\begin{equation}
  \mathcal{F} \left(\vec{\nabla}^z y(\vec{r})\right) = -|k|^z \mathcal{F}(y(\vec{r})), \ z>0,
\end{equation}
where $y(\vec{r})$ is an inifnitely differentiable function. Hence, the generalized
continuity equation gives rise to the following Green’s function in momentum
space
\begin{equation}
  G(\vec{k})= \frac{1}{\bar{\gamma}|k|^z + \bar{g}\epsilon}.
\end{equation}
Uncorrelated disorder $\langle \delta g(\vec r)\delta g(\vec r') \rangle = (\delta g)^2 \delta(\vec r-\vec r')$ gives the following scaling of temperature fluctuations with the dimension $d$ and fractional power $z$ 
\begin{align}
\ave{\delta T^2}
&=\frac{\epsilon^2\theta^2}{L}\int G(\vec{r}-\vec{r'})G(\vec{r}-\vec{r''})
\langle\delta g(\vec{r'})\delta g(\vec{r''}) \rangle d\vec{r}d\vec{r'}d\vec{r''}\notag \\
&=\frac{\epsilon^2\theta^2\delta g^2}{(2\pi)^d}\int d\vec{k} \frac{1}{(\bar{\gamma} |k|^z + \epsilon \bar{g})^2}\notag \\
&\sim \frac{\theta^2\delta g^2}{\bar{g}^2} \frac{\bar{g}^{d/z}}{\bar\gamma^{d/z}} \, \frac{(z-d)}{z^2 \sin(d\pi/z)} \, \epsilon^{d/z} ,
\end{align}
which is the result quoted in Eq.~(2).
\vspace{1cm}

\subsection*{Diffusive behavior in two dimensions}

In this section we generalize the random resistor model, used to describe the sub-diffusive behavior in the Griffiths regime of one dimensional systems, to the case of a two dimensional system. We will show how diffusive behavior emerges in this model throughout the thermal phase as long as the correlation length associated with the transition remains finite.

\begin{figure}[t!]
\center
\includegraphics[width=.9\linewidth]{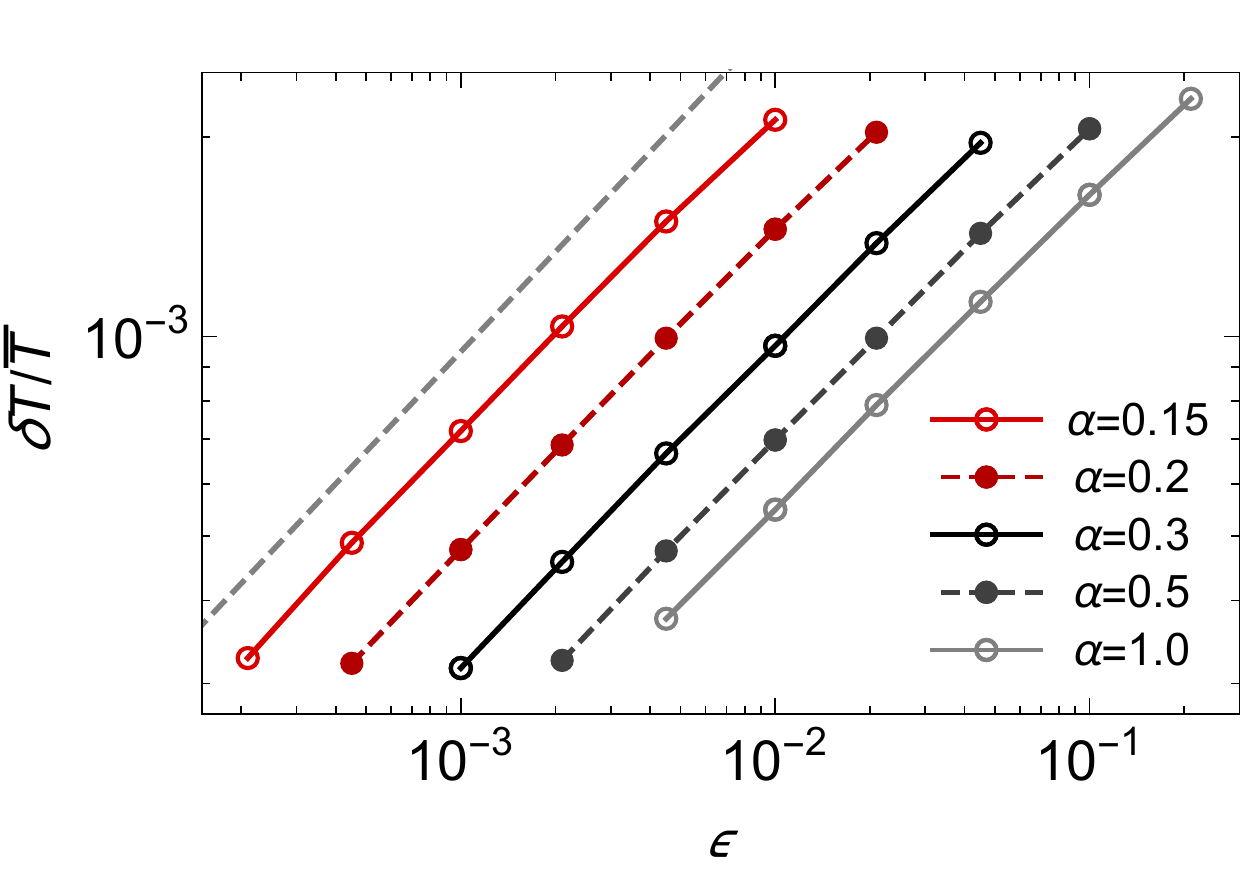}
\caption{Fluctuations of the local temperatures, $\delta T/\bar{T}$ as function of the coupling strength $\epsilon$ to non-thermal baths, computed from a two-dimensional resistor network model with conductances \eqref{EqGammaD} distributed as in Eq.~\eqref{EqPGammaD}.
For $\epsilon\to 0$, a diffusive behavior with temperature fluctuations described by
$\delta T/\bar{T} \sim \epsilon^{d/2z}$ with $z=2$ is observed for all $\alpha$.
The gray dashed line shows $\delta T/\bar{T} \sim \epsilon^{1/2}$ for comparison. 
Parameters:  $\kappa_0=1$, $a\Gamma_0=5.0$, $T_0=1$, $\delta g_1=0.05$, $\delta
g_2=0.05$, $\theta = T_0$ $N\times N$ system with $N=100$. Averaging over 500 configurations is performed.
\label{FigRN2D}
}
\end{figure}

Our two dimensional toy model consists of a square lattice of conducting islands connected by insulating links of linear length $\ell$ taken from a probability distribution  $p(\ell)\sim e^{-(\ell/\xi)^2}$ (and generally in $d$ dimensions $p(\ell)\sim e^{-(\ell/\xi)^d}$). This takes into account that in higher dimensions insulating regions are harder to construct than in one dimension because they can be short circuited by conducting paths. Conductances across the links are given by 
\begin{equation}\label{EqGammaD}
\Gamma = \epsilon\frac{\kappa_0}{\ell^{2-d}} + \Gamma_0 e^{-\ell/a}, \quad 
\alpha=\frac{a}{\xi}.
\end{equation}
This relation between the conductance and the link size implies a probability distribution of conductances that decays faster than any power law:
 \begin{align}\label{EqPGammaD}
P(\Gamma) \sim \frac{1}{\Gamma}
e^{-\alpha^d \left( \ln\left( \frac{\Gamma_0}{\Gamma} \right) \right)^d}\qquad \text{for } \ \Gamma \gg \epsilon\frac{\kappa_0}{a^{2-d}}.
\end{align}
This contrasts with the power-law distribution obtained in one dimension 
$P(\Gamma)\sim \left({\Gamma_0}/{\Gamma}\right)^{1-\alpha}$.

Fig.~\ref{FigRN2D} shows temperature fluctuations as a function of coupling strength to the environment $\epsilon$ for $d=2$. For all values of $\xi$ we see a dependence $\delta\beta/\bar{\beta}\sim \epsilon^{1/2}$ consistent with diffusive behavior $z=2$. This is in contrast to one dimension where we observed a continuously varying dynamical exponent $z\sim \xi/a$. This observation confirms the expectation \cite{gopalakrishnan16a} that Griffiths effect are absent in $d>1$ where insulating regions cannot serve as bottlenecks, but are rather short-circuited by surrounding smaller resistivities.

\subsection*{Current dependence on the dynamical exponent}

As an alternative to temperature fluctuations, we proposed in the main text, a setup that is driven via a small bias at the edges. In the bulk, the system is still coupled to phonons but not necessarily driven. The information about the dynamical exponent is in this case contained in the dependence of the current on the strength of coupling to the thermal phonon bath. The average resistivity, obtained by taking an ensemble average of the local conductances, was $\bar{\rho}\sim \left(\epsilon/\ln\epsilon^{-1}\right)^{\alpha-1}$, where $\alpha=a/\xi \sim 1/z$.

\begin{figure}[t!]
\center
\includegraphics[width=.86\linewidth]{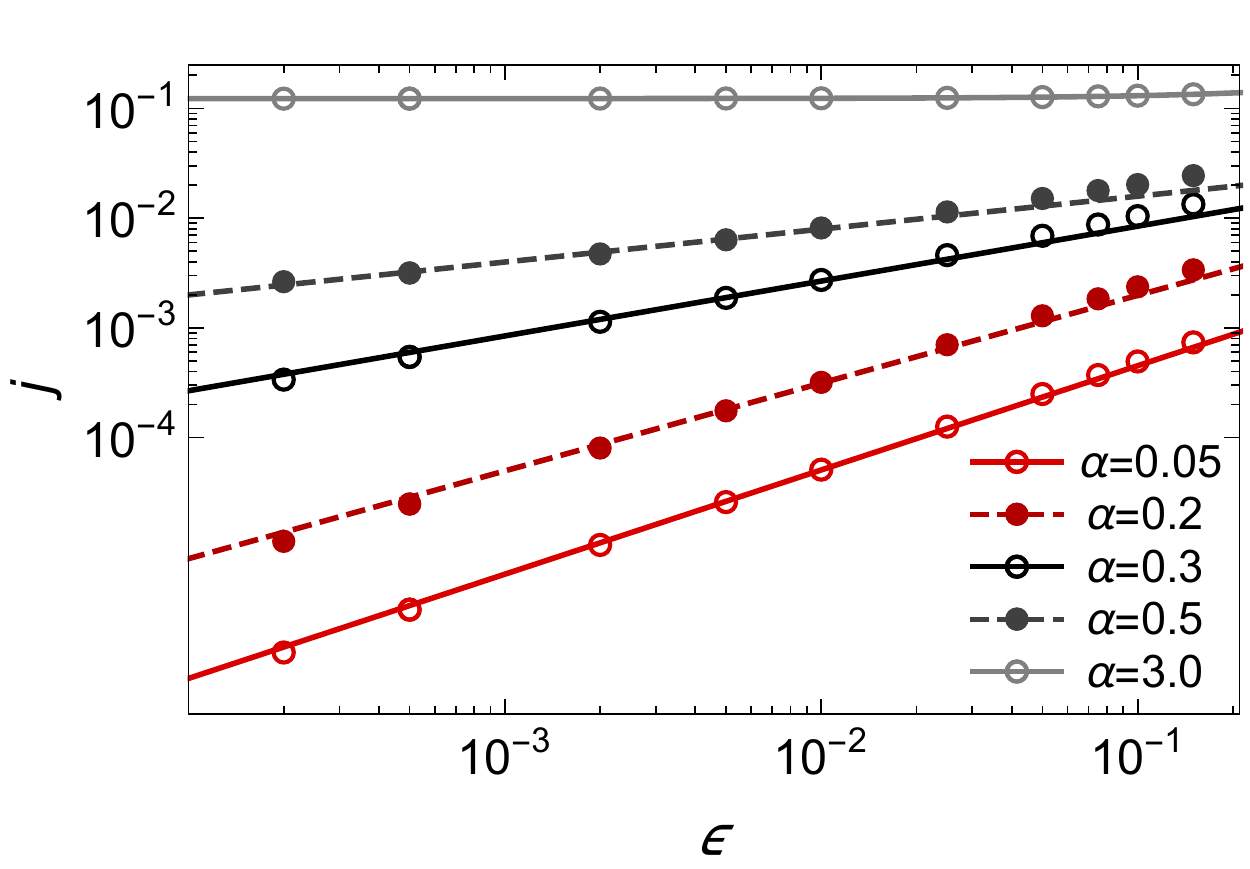}
\caption{Dependence of particle current $j$ on the strength of coupling to the baths $\epsilon$ in $d=1$. For $\alpha<1$, $j\sim\epsilon^{1-\alpha}$ is observed. For diffusive $\alpha >1$, on the other hand, a finite  $j(\epsilon\to 0)$ is observed. 
Parameters: $\sigma_0=1$, $a\tilde{\Gamma}_0=5.0$, $N=2000$, $V=0.1 N$
}\label{FigCurr}
\end{figure}

\begin{figure}[t!]
\center
\includegraphics[width=.84\linewidth]{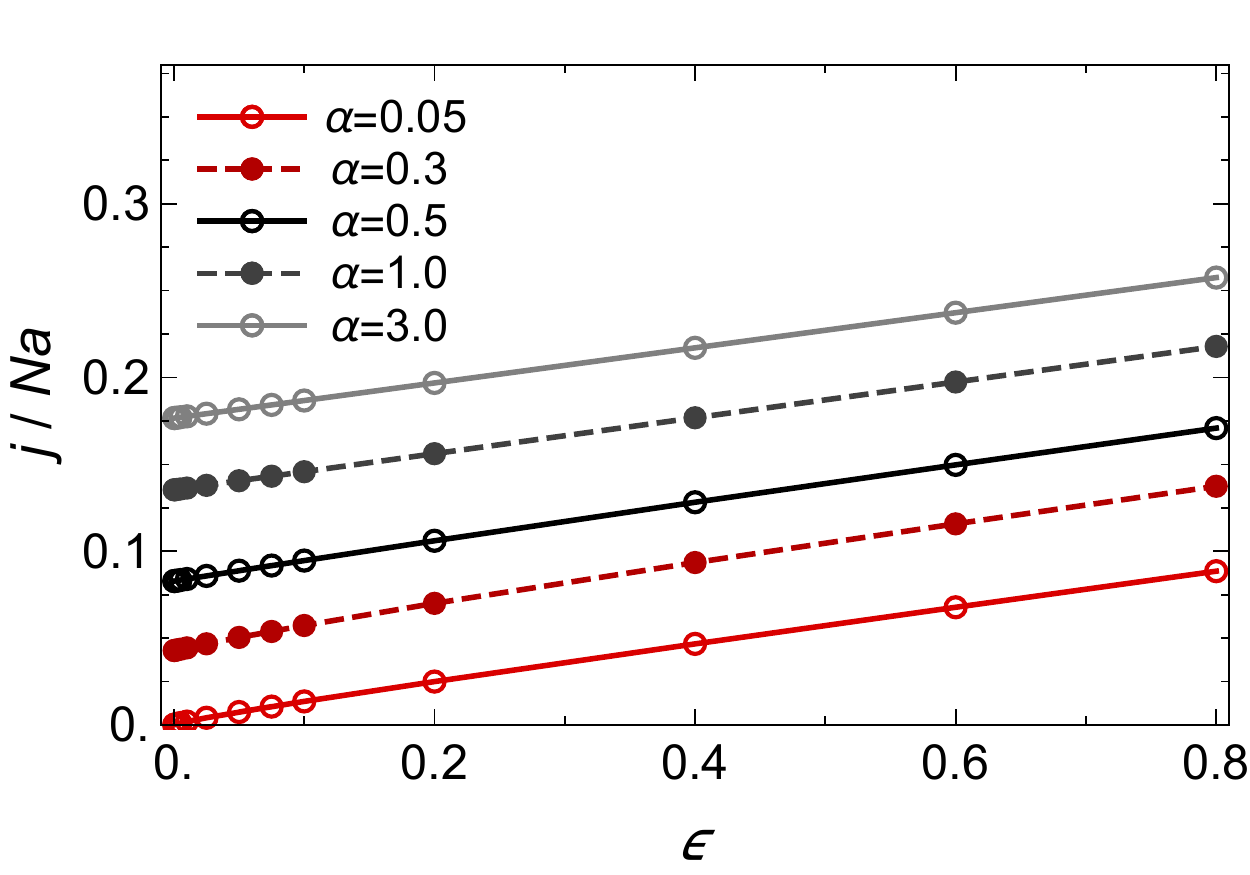}
\caption{ 
In $d=2$ dimension, current density $j/Na$ shows a similar behavior as the $d=1$ diffusive case $(\alpha>1)$ and can be, to a good approximation, fitted with $c_1 + \epsilon c_2$.
Parameters: $a\tilde{\Gamma}_0=5$, $\sigma_0=1$, system of $N\times N$ sites for $N=100$, $V=0.1 N$.
}\label{FigCurr2D}
\end{figure}

In this section we compute the resistance of the network in one and two dimensions, obtained through numerical  solution of the steady state rate equations. In the one dimensional case, for example, Kirchoff's law is
\begin{equation}\label{EqChargeNetwork}
 \big(\tilde{\Gamma}_{i,i+1} (\mu_{i+1}-\mu_i) -  \tilde{\Gamma}_{i-1,i} (\mu_{i}-\mu_{i-1})\big)=0.
\end{equation} 
In this case we set up a voltage bias across the chain $V=\mu_0-\mu_N$, then solve for the island chemical potentials $\mu_i$ to get the current. As in the thermal case, we draw the link sizes along the chain from the distribution $p(\ell)\sim e^{-(\ell/\xi)^d}$, while the link conductances are:
\be
\tilde{\Gamma}= \epsilon \frac{\sigma_0}{\ell^{2-d}} + \tilde{\Gamma}_0 e^{-\ell/a}
\ee
In the two dimensional case we set up a constant chemical potential $\mu=V$ on the left edge and a constant $\mu=0$ on the right edge.

The result of the calculation of the current in a one dimensional chain for the case $\alpha <1$ is shown in
Fig.~\ref{FigCurr}. The result agrees well with the analytic prediction for the resistance scaling as $\epsilon^{\alpha-1}$, which was obtained in the main text up to logarithmic corrections. 
In the diffusive regime $(\alpha>1)$, the current has a non-zero $\epsilon\to 0$ limit, as we expect for a system with finite intrinsic resistivity.
Also in two dimension we always find a finite resistivity in the $\epsilon\to 0$ as well as linear corrections, as shown in Fig.~\ref{FigCurr2D}. 

Remember that the dynamical exponent is related to $\alpha$ as $z\sim\alpha^{-1}$. Measuring the dependence $j(\epsilon,\alpha)$ in a disordered system would therefore provide information on $z$ and it's divergence upon approaching the MBL transition even in disordered materials that are weakly coupling to phonons. The strength of coupling to phonons $\epsilon$ can be tuned by controlling the phonon temperature; see our previous work \cite{lenarcic18}, where we showed that phonon temperature determines the effective coupling. 


\subsection*{Comparison to standard measures of dynamical scaling}
\label{sec:spread}
\begin{figure}[b!]
\center
\includegraphics[width=.84\linewidth]{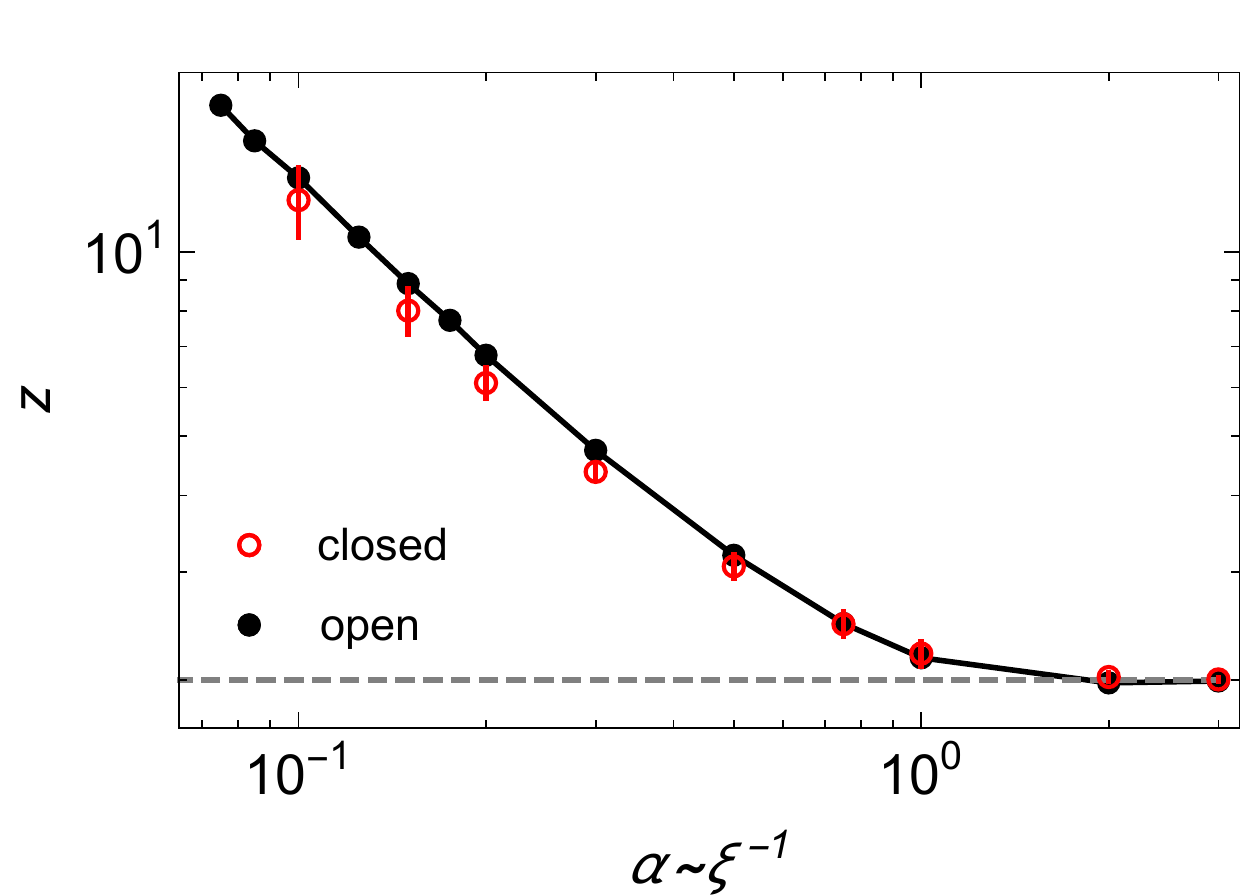}
\caption{Comparison of the dynamical exponent $z$ calculated from the fluctuations of the local temperatures (open) or from the from anomalous heat diffusion in a closed system (closed). 
Parameters: $\kappa_0=1$, $a\Gamma_0=5.0$, $T_0=1$, $\delta g_1=0.05$, $\delta g_2=0.05$, $\theta=T_0$, $N=1000$, averaged over $M=500$ configurations.
\label{FigRNSpread}
}
\end{figure}

In this section we verify that the dynamical exponent $z$ obtained from the temperature variations, through the relation $\delta T \sim \epsilon^{d/2z}$ is identical to the standard measure of dynamical scaling obtained from spreading of  an energy fluctuation. To obtain the standard measure of $z$ we consider an initially localized energy profile in the random resistor network $e_i = \delta_{i,x_0}$. We determine $z$ by measuring the width of the energy packet $\sigma_E=\sqrt{\sum_i (i-x_0)^2 \tilde{e}_i(t)} $ and fitting it to a power law  $\sigma_E\sim t^{1/z}$.  Here $\tilde{e}_i(t) =  \langle\!\langle e_i(t) \rangle\!\rangle $ is a disorder averaged energy distribution. For each realization of the disorder  $e_i(t)$ is calculated using the continuity equation with $\epsilon=0$:
\begin{equation}
\partial_t e_i- \Gamma_{i,i+1} (T_{i+1}-T_i) +  \Gamma_{i-1,i} (T_{i}-T_{i-1})=0
\end{equation}
Note that here $\Gamma = \Gamma_0 e^{-\ell/a}$. Fig.~\ref{FigRNSpread} shows a qualitative agreement between the value of $z$ obtained via these two different approaches.


\subsection*{Convergence in system size, time and bond dimension}
\label{sec:beta-dynamics}

\begin{figure}[b!]
  \centering
 \includegraphics[width=.9\linewidth]{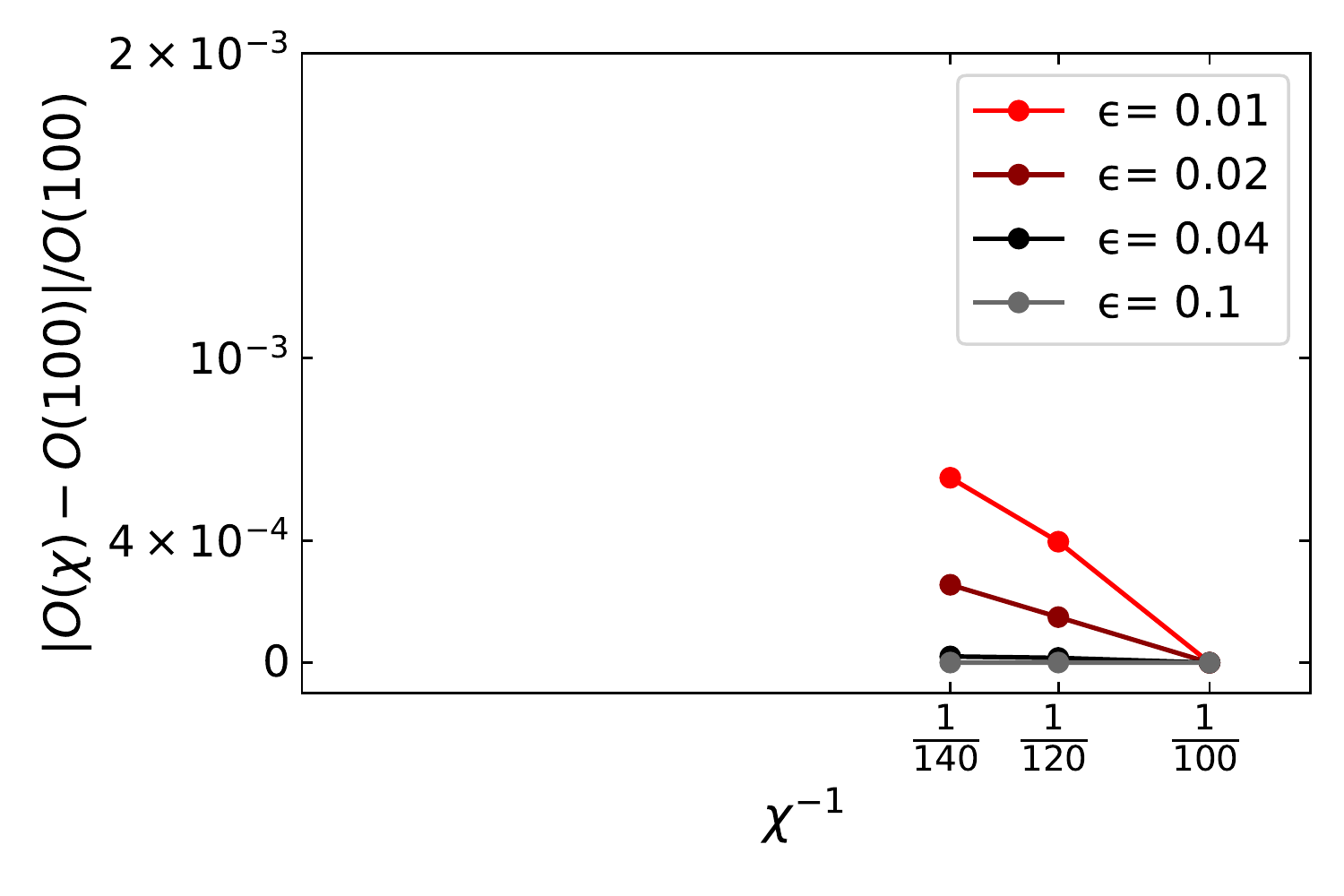} 
  \caption{Relative error due to finite bond dimension $\chi=100$ can be estimated from the ratio 
  $|O(\chi)-O(\chi=100)|/O(\chi=100)$, 
  $O(\chi)=\frac{\delta \beta}{\bar{\beta}}(\chi)$. The error estimated from the $\chi\to\infty$ extrapolation is below 0.1\% for smallest $\epsilon=0.01$ used in our computations. Parameters: $N=20$, $h=4$, with averaging over $60$ realizations.
  }
  \label{fig:chi-dependence}
\end{figure}

In this section, we look more closely into what are the limiting factors of the calculation.
We first investigate how our results on system size $N=20$ depend on the bond dimension. Fig.~\ref{fig:chi-dependence} shows the relative change of the expectation value of $O=\delta\beta/\beta(\epsilon)$ with bond dimension $\chi$ at steady state. We set $\chi=100$, used for the results in the main text, as a baseline.  $O(\chi)$ is averaged over 60 disorder realizations, which are
the same for different $\chi$. We find that the error extrapolated to $\chi =\infty$ is small, e.g., for $h=4$ below 0.1\%. We also
note that the extrapolated error estimate grows with decreasing $\epsilon$,
hence for smaller values of $\epsilon$ a larger bond dimension will be required, increasing the costs of computations at small $\epsilon$.

\begin{figure}[b!]
  \centering
\includegraphics[width=.85\linewidth]{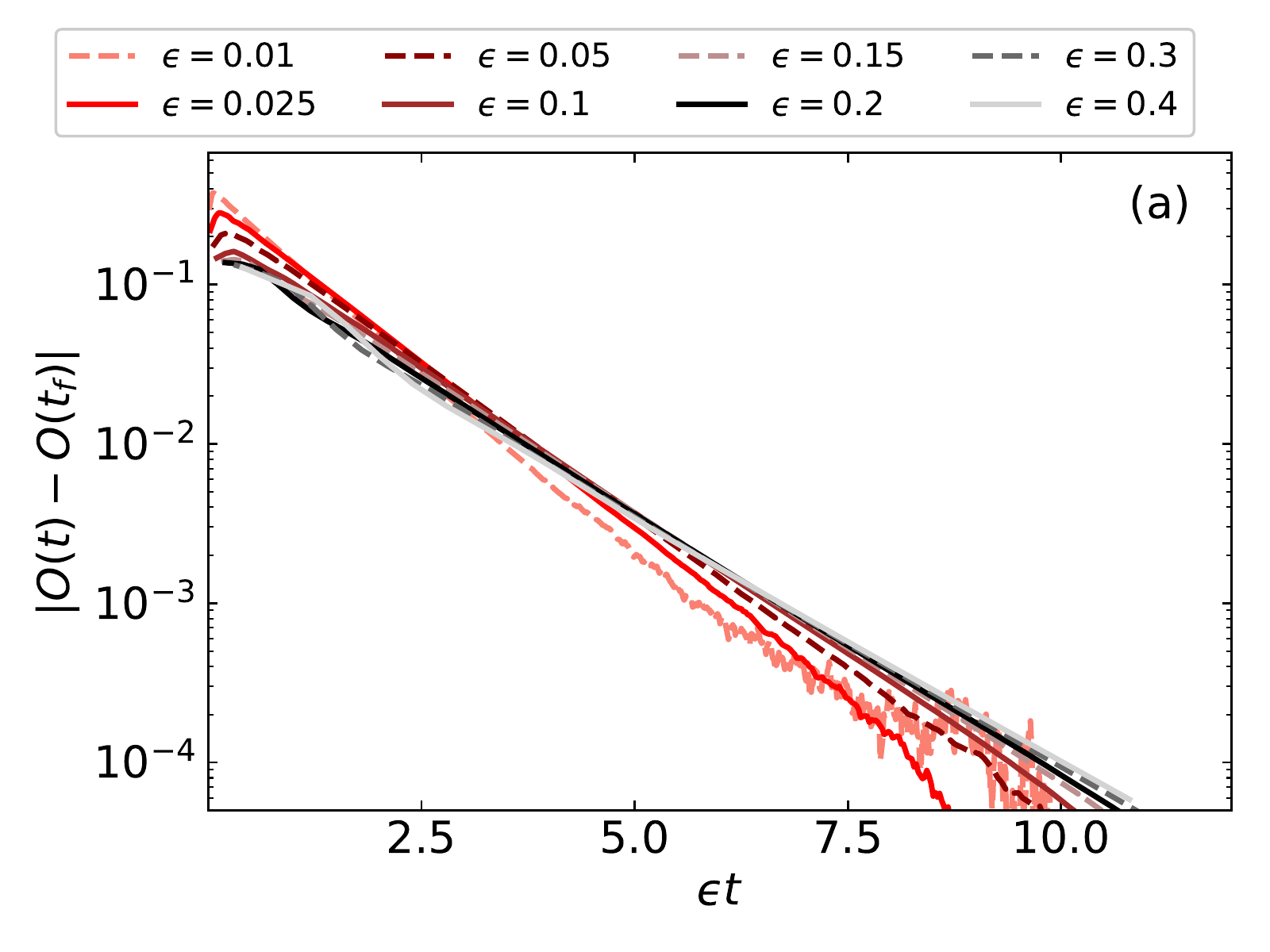} 
\includegraphics[width=.83\linewidth]{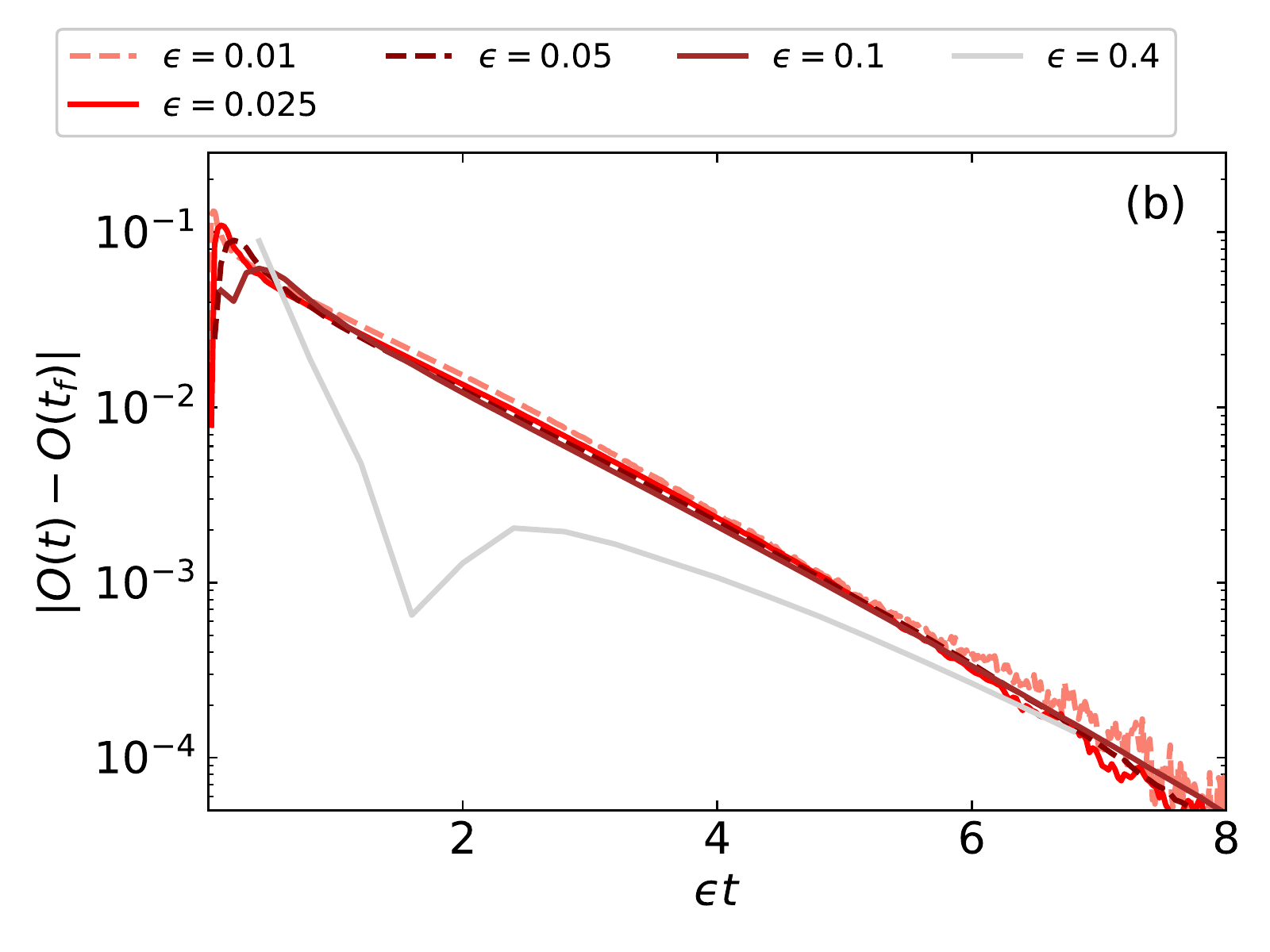} 
  \caption{Evolution of $O(t)=\frac{\delta \beta}{\bar{\beta}}(t)$ (with respect to $O(t_f)$ at maximal propagation time $t_f$) during the TEBD computation,
  for one disorder realization at $h=4$, $\chi=100$, for systems of size (a) $N=20$ and (b) $N=40$. Note that the 
  time axis is rescaled by $\epsilon$ to reveal an exponential relaxation with a convergence rate proportional to $\epsilon$.}
  \label{fig:convergence-rate-beta}
\end{figure}

In Fig.~\ref{fig:convergence-rate-beta}  we show the relaxation of temperature fluctuations $O(t)=\frac{\delta \beta}{\bar{\beta}}(t)$ to steady state in the TEBD time evolution. Specifically, we plot $|O(t)-O(t_f)|$ with respect to the $O(t_f)$ at the maximal propagation time $t_f$. As expected, we see exponential relaxation to the
steady-state value with a characteristic rate which scales as $\epsilon$. Obtaining results for smaller $\epsilon$ is thus increasingly hard with TEBD.

Assuming that the necessary bond dimension $\chi$ is independent of system size,
our approach should be scalable, with computational demands growing linearly
with system size. Figs.~\ref{fig:convergence-rate-beta}(a,b) compare the
convergence of $O(t)=\frac{\delta \beta}{\bar{\beta}}(t)$ for a single
realization at $N=20,40$ system sizes. While the computational time
approximately doubles, the exponential convergence rate with TEBD evolution time
$t$ is comparable.

\begin{figure}[t!]
  \centering
 \includegraphics[width=0.85\linewidth]{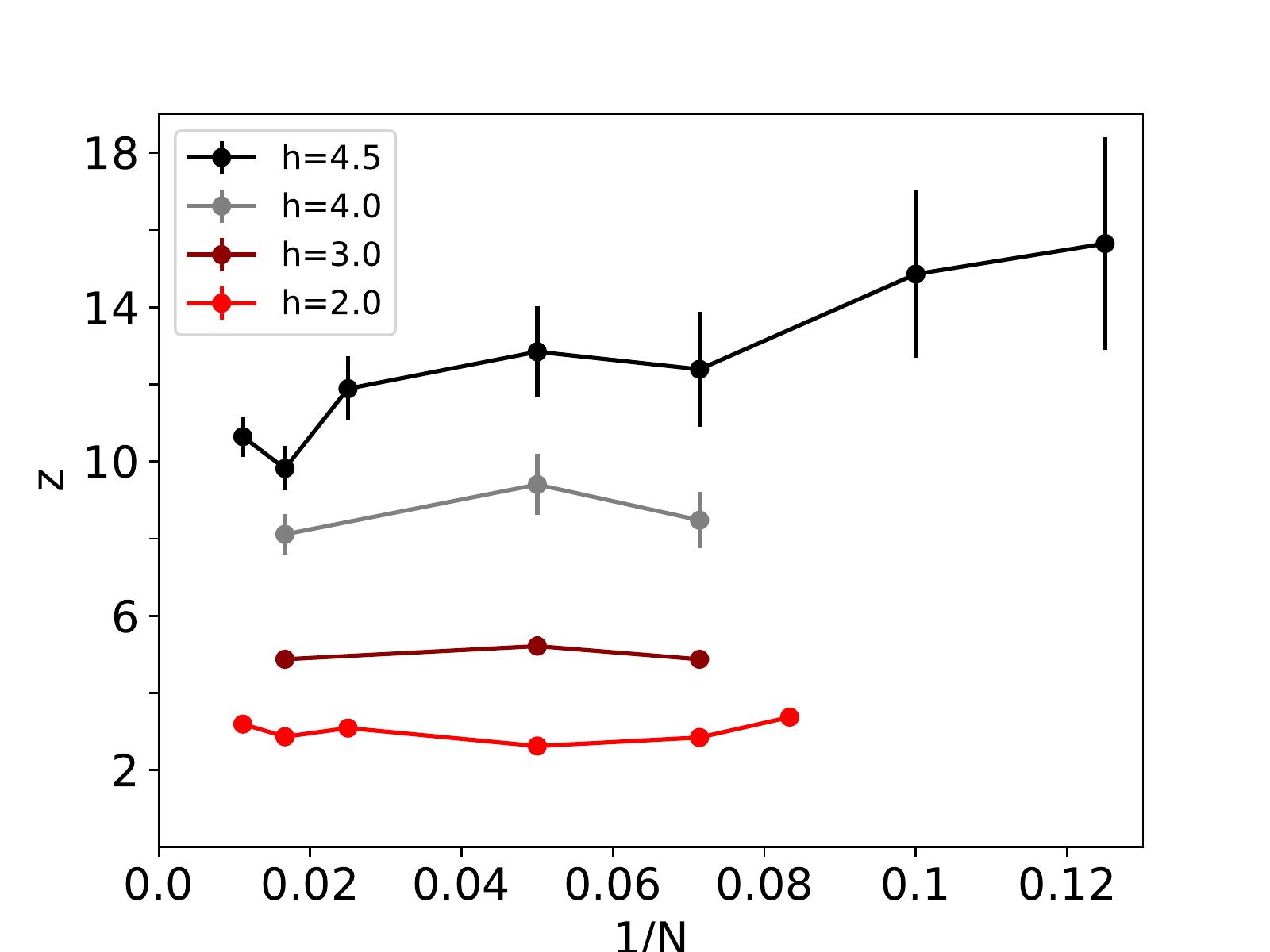}
  \caption{Finite-size analysis of the dynamical exponent $z$ for $h=2.0,3.0,4.0,4.5$. Calculations are performed with bond dimension $\chi=100$ and $\epsilon\geq 0.01$. A different number of realizations is used for different points, e.g., at $h=2.0$ and $N\in[12,90]$, $150-50$ realizations are used, for $h=4.5$ and $N\in[8,90]$, $790-190$ realizations are used. Fit error bars are obtained with the jackknife resampling of data with $\epsilon\geq 0.01$ and do not estimate systematic deviations from the $z$ that would be obtained using $\epsilon<0.01$, necessary for $h>4$.}
  \label{fig:z-finite-size}
\end{figure}

Finally, we present a finite size scaling analysis of the results. We show the dependence of the dynamical exponent $z$, obtained from our numerical scheme, on the system size. The exponent $z$ is extracted by fitting a power law for the dependence of the temperature variance on $\epsilon$ with $\epsilon\ge 0.01$, Eq.(9) in the main text. The results are shown in Fig.~\ref{fig:z-finite-size}. 

In systems with disorder strengths $h=2.0,3.0,4.0$ we do observe negligible finite size dependence. Recall that this is the range of $h$ we used to extract the power law divergence of $z\sim (h_c-h)^{-\nu}$ on approaching the critical point. It is encouraging to see that this behavior is unaffected by finite size.  We do see a non systematic finite size dependence for disorder strength $h=4.5$, which we attribute to uncertainty in fitting the dynamical exponent $z$. Indeed, as noted in the main text, for $h>4$ ($z>8$) we can no longer extract a reliable power law fit to $\frac{\delta \beta}{\bar{\beta}}\sim \epsilon^{1\over 2z}$ in the range $\epsilon\ge 0.01$. In order to reliably obtain larger values of $z$ close to the critical point  one would have to reduce the cutoff $\epsilon$ exponentially in $z$ (equivalently, in $\xi$). Thus it is the finite time cutoff ($1/\epsilon$) rather than the finite size, which limits the calculation.

As noted in the main text, calculations performed at smaller $\epsilon$ might yield somewhat larger $h_c$ as well, which would, in turn, impact the value of $\nu$ obtained from $z\sim(h_c-h)^{-\nu}$ fit.

\new{
\subsection*{Hydrodynamic equations as an expansion of Lindblad driving}
}
In the main text we introduced the hydrodynamic approach  as an effective description of a system that is coupled to a thermal (e.g. phonon) bath and to a drive (e.g. white light). Here we make the connection to the microscopic calculation, that was performed for a spin chain coupled to Markovian nonequilibrium baths described by Lindblad operators. We show that the hydrodynamic Eq.~(1) can be derived using an expansion in small temperature variations around the thermal density matrix, determined from the Liouville equation $\dot{\rho}=(\cl{L}_0 + \epsilon \cl{D})\rho=0$, where $\cl{L}_0\rho=-i[H,\rho]$ and $\cl{D}$ corresponds to the dissipator super-operator.

On the ergodic side, 
the system approaches a thermal state for $\epsilon\to 0$ \cite{lenarcic18}. For small epsilon, we can therefore expand the steady state density matrix in weak temperature variations around the thermal state
\begin{equation}\label{EqExpand}
\rho\approx\rho_0(\bar{T}) + 
\sum_j\delta T_j\frac{\partial\rho}{\partial T_j}\Big|_{T_j=\bar{T}}  + \cdots , \
\rho_0(\bar{T})\equiv
\frac{e^{-H/\bar{T}}}{\tr[e^{-H/\bar{T}}]}
\end{equation}
We will now use the expansion \eqref{EqExpand}  in order to show how the phenomenological terms in Eq.~(1) can emerge from the microscopic Liouville equation.

First of all, the term $-\epsilon \bar{g} (T(\vec{r})-\bar{T}))$ ensures the relaxation towards the correct mean temperature $\bar{T}$, which is determined from the stationarity condition applied to the total rate equation for the energy \cite{lenarcic18},
\begin{align}\label{EqEn}
\ave{\dot{H}}=
\tr[H(\cl{L}_0 + \epsilon\cl{D})\rho_0(\bar\beta)]
=\tr[H \, \epsilon\cl{D} \, \rho_0(\bar\beta)]
\stackrel{!}{=}0
\end{align}

To see the emergence of the other terms in Eq.~(1) we consider the behaviour of the local energy density $\ave{h_i}$, where $H=\sum_i h_i$,
\begin{align}
\frac{d}{dt}\ave{h_i}
=&\tr\left[h_i \cl{L}_0 \rho\right] \label{EqDivCurr} \\
&+\tr\left[h_i \epsilon \cl{D} \rho_0(\bar\beta)\right]\label{Eqg2}\\
&+\tr\left[h_i \epsilon \cl{D} \sum_j\frac{\partial\rho}{\partial T_j}\Big|_{T_j=\bar{T}}\delta T_j\right] \label{Eqg1}\\
&+ \cdots \notag 
\end{align}
Using the definition for the energy currents,
$j_{i,i+1}=i[h_i,h_{i+1}]$,
we can see that the right hand side of expression \eqref{EqDivCurr} equals to the difference in expectation value of energy currents across neighboring links. On the other hand, in a system with spatially varying local temperatures, local current expectation values are proportional to local temperature gradients
\begin{align}
\tr\left[h_i \cl{L}_0 \rho\right]
&=\tr[(-j_{i,i+1}+j_{i-1,i})\rho]\\
&=\Gamma_{i,i+1} (T_{i+1}-T_i) -  \Gamma_{i-1,i} (T_{i}-T_{i-1})\notag\\
&\sim \nabla \cdot(\kappa(\vec{r}) \nabla T(\vec{r})) 
\end{align}

The term \eqref{Eqg2} corresponds to the gain and loss of local energy density due to the driving and dissipation, evaluated with respect to the homogeneous thermal state 
\begin{equation}
\tr\left[h_i \epsilon \cl{D} \rho_0(\bar\beta)\right]
\sim \epsilon \, \theta g_2(\vec{r}).
\end{equation}
Here 
$\ave{g_2(\vec{r_i})}
=\frac{1}{\theta N}\sum_{i}\tr[h_i\cl{D} \rho_0(\bar\beta)]
=0$ due to Eq.~\eqref{EqEn}.

Term \eqref{Eqg1} is of the same type, but comes from the next order expansion in the variation of local temperatures
\begin{equation}
\tr\left[h_i \epsilon \cl{D} \sum_j\frac{\partial\rho}{\partial T_j}\Big|_{T_j=\bar{T}}\delta T_j\right]
\sim \epsilon g_1(\vec{r}) \, \delta T (\vec{r})
\end{equation}

Collecting the dominant terms in the $(\delta T (\vec{r}))^n$ expansion, we get the hydrodynamic relation 
\begin{equation}\label{EqContinuity}
\partial_t e - \nabla \cdot(\kappa(\vec{r}) \, \nabla T(\vec r)) = - \epsilon\,  g_{1}(\vec r) (T(\vec r)-\bar{T}) + \epsilon\,\theta g_2(\vec r) 
\end{equation}
which can be identified with Eq. (1), except that $T_0$ is replaced by $\bar{T}$ (set by Eq.~\ref{EqEn}) and that the last term contains only the random part with a zero mean, i.e., $g_2(\vec{r})=\delta g_2(\vec{r})$.

\bibliographystyle{physrev4}
\bibliography{MBL,weakly_open_library_all}

\end{document}